\documentclass[twocolumn,pra,aps,superscriptaddress]{revtex4}
\usepackage[T1]{fontenc}
\usepackage[latin9]{inputenc}
\usepackage{amsmath}
\usepackage{amssymb}
\usepackage{graphicx}
\usepackage{esint}

\usepackage[english]{babel}
\usepackage{times}
\usepackage{booktabs}
\usepackage{array}
\usepackage{supertabular}
\usepackage{multirow}
\usepackage{calc}
\usepackage{mathtools}
\usepackage{bm}
\usepackage{stackrel}
\usepackage{setspace}
\usepackage{subfigure}
\usepackage{indentfirst}
\usepackage{cases}
\usepackage[colorlinks,linkcolor=blue,citecolor=blue,hyperindex,bookmarks=false,pdfstartview=FitH,dvipdfm]{hyperref}
\usepackage{bm}



\begin{document}

\title{Enantio-specific state transfer for symmetric-top chiral molecules}

\author{Bo Liu}
\affiliation{Beijing Computational Science Research Center, Beijing 100193, China}

\author{Chong Ye}
\affiliation{Beijing Key Laboratory of Nanophotonics and Ultrafine Optoelectronic Systems School of Physics, Beijing Institute of Technology, Beijing 100081, China}
\affiliation{Beijing Computational Science Research Center, Beijing 100193, China}

\author{C. P. Sun}
\affiliation{Beijing Computational Science Research Center, Beijing 100193, China}
\affiliation{Graduate School of China Academy of Engineering Physics, No. 10 Xibeiwang East Road, Haidian District, Beijing 100193, China}

\author{Yong Li}
\email{liyong@csrc.ac.cn}
\affiliation{Beijing Computational Science Research Center, Beijing 100193, China}

\begin{abstract}
We study the enantio-specific state transfer in a four-level model for symmetric-top chiral molecules. Such a model is formed by coupling the electric-dipole transitions among four appropriate working states with three electromagnetic fields. It includes two cyclic three-level substructures, where the overall phases differ by $\pi$ with enantiomers and reflect the chirality dependence of the molecule. Based on this four-level model, two dynamic ways are proposed to achieve the approximately perfect enantio-specific state transfer for symmetric-top chiral molecules.
\end{abstract}

\date{\today}

\maketitle

\section{Introduction} \label{Introduction}
A chiral molecule is not superposable on its mirror image through pure translation or/and rotation. The left- and right-handed chiral molecules (called ``enantiomers'') coexist in many biologically active compounds and may have significant differences in physiological effects, pharmacological effects, and biological processes~\cite{Latscha,Hutt,Gal}. Only one enantiomeric form is biologically beneficial, while the other one may be harmful or fatal. Thus, the enantio-detection, enantio-separation, and enantio-conversion of chiral molecules are important and challenging tasks~\cite{Barcellona,Barcellona2,Gershnabel,Tutunnikov,Forbes,Yachmenev,Yurchenko,Cameron,Bradshaw,Zhang}.

In the past decades, the research of chiral molecules based on a cyclic three-level configuration~\cite{Shapiro,Salam,Fujimuraa,Hirota,Lehmann4,Jia1,Ye2,Chen,Xu,Kang,Chen2,Li2,ShapiroLiX,Kral2,Kral,Li3,Jia2,Ye1,Torosov1,Torosov2,Wu} has become remarkable in the physics of atomic, molecular, and optical physics.
In natural atoms, such a cyclic three-level system is forbidden, because three electric-dipole transitions can not coexist due to the electric-dipole selection rules. However, the cyclic three-level system can exist in chiral molecules and other symmetry-broken systems~\cite{Liu,WangZH2,Zhou}. 
Due to the intrinsic property of chiral molecules, the product of the corresponding three electric dipoles of the enantiomer in the cyclic three-level model differs in sign. So the overall phase of the product of three coupling strengths in the cyclic three-level model of the enantiomer differs by $\pi$. The chirality dependence of the overall phase makes the dynamics of the enantiomers different, then the enantio-specific state transfer~\cite{Kral2,Kral,Li3,Jia2,Ye1,Torosov1,Torosov2,Wu}, enantio-detection~\cite{Hirota,Lehmann4,Jia1,Ye2,Chen,Xu,Kang,Chen2}, and enantio-separation~\cite{Li2,ShapiroLiX} of chiral molecules can be achieved.

In the schemes of enantio-specific state transfer (as well as enantio-detection and enantio-separation) based on cyclic three-level systems~\cite{Shapiro,Salam,Fujimuraa,Hirota,Lehmann4,Jia1,Ye2,Chen,Xu,Kang,Chen2,Li2,ShapiroLiX,Kral2,Kral,Li3,Jia2,Ye1,Torosov1,Torosov2,Wu}, the molecular rotations usually have not been considered. For real gaseous molecules, the molecular rotations should be involved. In this case,
the ideal cyclic three-level model for asymmetric-top chiral molecules can be formed by applying three appropriate electromagnetic fields~\cite{Ye,Koch},
but it is impossible to realize such a model for symmetric-top chiral molecules due to the electric-dipole selection rules between their rotational states~\cite{Hornberger}.

In this paper, we study the enantio-specific state transfer in a four-level model for symmetric-top chiral molecules. Based on electric-dipole transitions, we select appropriately four working states and corresponding three electromagnetic fields to form the four-level model including two cyclic three-level substructures. In each cyclic three-level substructure, the product of the corresponding three electric dipoles changes sign with enantiomers. That is, the overall phase in each cyclic three-level substructure differs by $\pi$ with enantiomers, which reflects the chirality dependence of this model and indicates different dynamics for the enantiomers. Under the large-detuning condition, we further reduce the four-level model to an effective two-level one with the same detuning but different effective couplings for the enantiomers. We also use two dynamic ways to realize the approximately perfect enantio-specific state transfer for symmetric-top chiral molecules. Thereby, by a variety of energy-dependent processes, the enantiopure molecules can be further spatially separated from the initial chiral mixture~\cite{Li2,ShapiroLiX,Drewsen}.

The structure of the article is as follows.
In Sec.~\ref{How}, we introduce the electric-dipole ro-vibrational transition of symmetric-top chiral molecules.
In Sec.~\ref{K}, the four-level model with two cyclic three-level substructures is realized for symmetric-top chiral molecules with applying three electromagnetic fields.
In Sec.~\ref{D2S2}, taking the molecule $\text{D}_{2}\text{S}_{2}$ as an example, we show how to choose the vibrational wave functions of the working states.
In Sec.~\ref{V}, based on the four-level model, two dynamic ways are used to achieve the enantio-specific state transfer for symmetric-top chiral molecules.
Conclusions are given in Sec.~\ref{Conclusion}.

\section{Electric-dipole ro-vibrational transition of symmetric-top chiral molecules}\label{How}
The rotational Hamiltonian for a chiral molecule reads~\cite{Zare}
\begin{equation}
\hat{H}_{\text{rot}}=A \hat{J}_{z}^{2}+B \hat{J}_{x}^{2}+C \hat{J}_{y}^{2}
\label{eq:Hrot}
\end{equation}
with three rotational constants $A$, $B$, and $C$. As follows, we focus on the prolate symmetric-top chiral molecules ($A>B=C$). Here, $\hat{J}_{x,y,z}$ are angular momentum operators along the principal axes of the moment of inertia, respectively. The eigenstates of the rotational Hamiltonian~(\ref{eq:Hrot}) are $\left|J,K,M\right\rangle$ with the total angular momentum number $J$, the quantum number $K\,(-J \leq K \leq J)$, and the magnetic quantum number $M\,(-J \leq M \leq J)$. Here $K$ and $M$ are the projections of the angular momentum on the molecule-fixed $z$-axis and the space-fixed $Z$-axis, respectively. The corresponding eigenenergies are $\varepsilon_{J,K}=C J (J+1)+(A-C)K^{2}$~\cite{Zare}.

In the following discussion, we suppose that the coupling among vibration, rotation, and electronic wave functions of the molecule can be neglected. To elucidate our scheme simply, we assume all the working states are in the electronic ground state, and omit it later. Then the full wave function of the chiral molecule can be described by the basis $\{\left|l\right\rangle=\left|v_{l}\right\rangle\otimes\left|J_{l},K_{l},M_{l}\right\rangle\}$ with $\left|v_{l}\right\rangle$ the vibrational wave functions.

On the other hand, an electromagnetic field can be written in the linear combination,
$\emph{\textbf{E}}^{s}=\text{Re}\left\{\sum_{\sigma=0,\pm 1}\bm{e}^{s}_{\sigma} E^{s}_{\sigma} e^{-\text{i} (\omega t+\varphi_{\sigma})} \right\}$.
The notation ``$s$'' indicates the space-fixed frame. $E^{s}_{\sigma}$, $\omega$, and $\varphi_{\sigma}$ are the field amplitude, the frequency, and the initial phase of the electromagnetic field, respectively. $\sigma$ indicates the helicity component of the electromagnetic field with $\bm{e}^{s}_{0}=\bm{e}^{s}_{Z}$ and $\bm{e}^{s}_{\pm1}=(\text{i}\bm{e}^{s}_{Y}\pm\bm{e}^{s}_{X})/\sqrt{2}$. $\bm{e}^{s}_{X}$, $\bm{e}^{s}_{Y}$, and $\bm{e}^{s}_{Z}$ correspond to the unit vectors of axes of the space-fixed frame.

For the electric-dipole-allowed transition between a lower level $\left|l\right\rangle$ and an upper level $\left|j\right\rangle$, the interaction Hamiltonian $\hat{V}^{s}=\bm{\hat{\mu}} \cdot \emph{\textbf{E}}^{s}$ is obviously written as $(\hbar=1)$ $\hat{V}^{s}=\Omega_{jl} e^{-\mathrm{i}\omega t}
\left|j\right\rangle \left\langle l\right|+\text{H.c.}\,(j>l)$~\cite{Hornberger,Ye}. Here $\bm{\hat{\mu}}$ is the electric-dipole operator, and
$\Omega_{jl}$ is the coupling strength
\begin{eqnarray}
\Omega_{jl} & = & \frac{1}{2} \sum_{\sigma=0,\pm 1} E^{s}_{\sigma} e^{-\mathrm{i}\varphi_{\sigma}} \left\langle j \right| \hat{\mu}^{s}_{\sigma} \left|l\right\rangle \label{eq:Omega}\\
& = & \frac{1}{2}\sum_{\sigma,\sigma^{\prime}=0,\pm 1} \sqrt{\left(2J_{j}+1\right)\left(2J_{l}+1\right)} \left\langle v_{j} \right| \hat{\mu}_{\sigma^{\prime}}^{m} \left|v_{l}\right\rangle E^{s}_{\sigma} \nonumber\\
&   & \times   \left(-1\right)^{\sigma+\sigma^{\prime}-K_{l}+M_{l}}e^{-\text{i}\varphi_{\sigma}} W^{(\sigma)}_{J_{j}M_{j},J_{l}M_{l}} W_{J_{j}K_{j},J_{l}K_{l}}^{(\sigma^{\prime})}.\nonumber
\end{eqnarray}
Here $\hat{\mu}^{s}_{\sigma}=\bm{\hat{\mu}}\cdot\bm{e}^{s}_{\sigma}$ and $\hat{\mu}^{m}_{\sigma^{\prime}}=\bm{\hat{\mu}}\cdot\bm{e}^{m}_{\sigma^{\prime}}$ are the components of the electric dipole in the space-fixed frame and the molecule-fixed frame, respectively. $\sigma^{\prime}$ indicates the spherical components of the electric dipole $\hat{\mu}^{m}_{\sigma^{\prime}}$ in the molecule-fixed frame with $\bm{e}^{m}_{0}=\bm{e}^{m}_{z}$ and $\bm{e}^{m}_{\pm1}=(\text{i}\bm{e}^{m}_{y}\pm\bm{e}^{m}_{x})/\sqrt{2}$. And $\bm{e}^{m}_{x,y,z}$ correspond to the unit vectors of axes of the molecule-fixed frame. Here the $3j$ symbol reads
\begin{eqnarray}
W_{JM,J^{\prime}M^{\prime}}^{(\sigma)}
&=&\left(\begin{array}{ccc}
      J & 1       & J^{\prime} \\
      M & -\sigma & -M^{\prime}
      \end{array}\right).
\end{eqnarray}

\begin{figure}[htbp]
\centering
\includegraphics[width=0.90\linewidth]{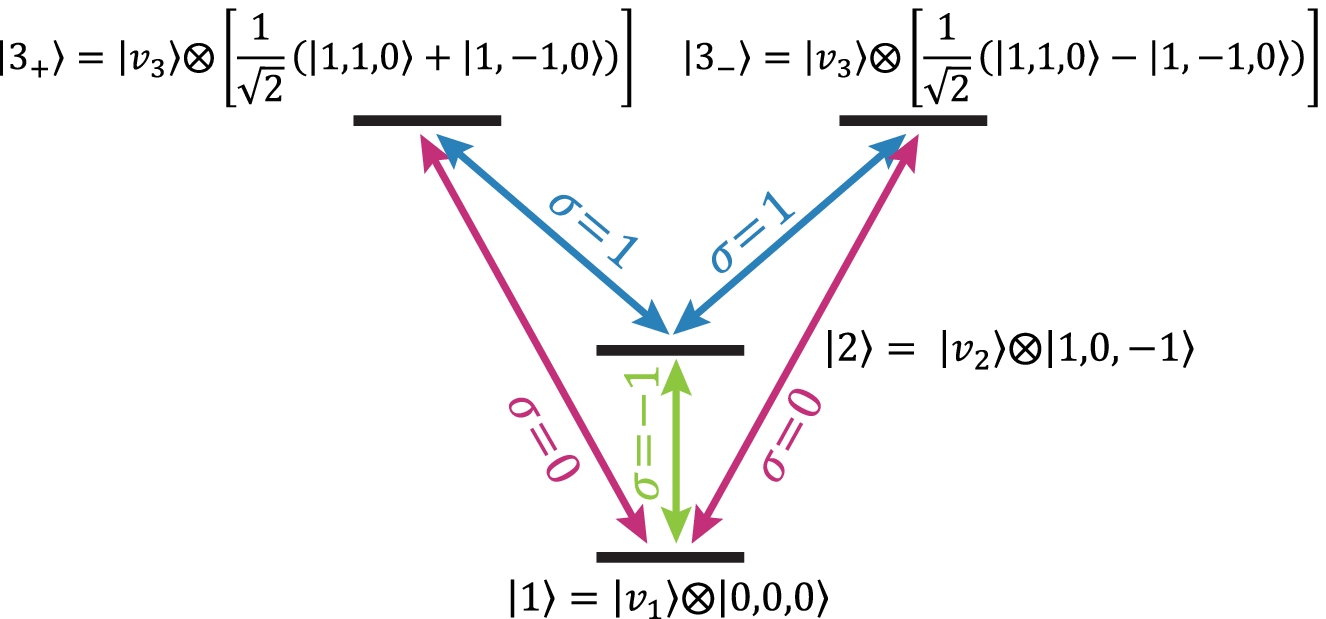}
\caption{The four-level model for the symmetric-top molecule. Two cyclic three-level substructures $\left|1\right\rangle \rightarrow \left|2\right\rangle \rightarrow \left|3_{+}\right\rangle \rightarrow \left|1\right\rangle$ and $\left|1\right\rangle \rightarrow \left|2\right\rangle \rightarrow \left|3_{-}\right\rangle \rightarrow \left|1\right\rangle$ coexist under the interaction with three electromagnetic fields. $\sigma=0$ corresponds to the linearly $Z$ polarized electromagnetic field with the polarization vector $\bm{e}^{s}_{0}=\bm{e}^{s}_{Z}$. $\sigma=1\,(\sigma=-1)$ corresponds to the circularly polarized electromagnetic field rotating about the $Z$ axis in the right-handed (left-handed) sense with the polarization vector $\bm{e}^{s}_{1}=(\text{i}\bm{e}^{s}_{Y}+\bm{e}^{s}_{X})/\sqrt{2}~[\bm{e}^{s}_{-1}=(\text{i}\bm{e}^{s}_{Y}-\bm{e}^{s}_{X})/\sqrt{2}]$.
\label{fig:model}}
\end{figure}

We are interested in the condition that the coupling strengths in Eq.~(\ref{eq:Omega}) are non-zero. Here, the $3j$ symbols play a key role in determining the electric-dipole selection rules, that is $\Delta J=J_{j}-J_{l}=0,\pm1$, $\Delta M=M_{j}-M_{l}=\sigma$, and $\Delta K=K_{j}-K_{l}=\sigma^{\prime}$~\cite{Hornberger}. These electric-dipole selection rules offer the possibility of forming the four-level model for symmetric-top chiral molecules.

\section{Chirality-dependent four-level configuration} \label{K}
In this section, we realize the four-level model for symmetric-top chiral molecules by choosing appropriate states $\left|J,K,M\right\rangle$ according to the electric-dipole selection rules. For simplicity, we focus on the subspace $J \leq 1$. The working states are selected as
\begin{eqnarray}
\left|1\right\rangle  & \equiv & \left|v_{1}\right\rangle\otimes\left|0,0,0\right\rangle,\nonumber \\
\left|2\right\rangle  & \equiv & \left|v_{2}\right\rangle\otimes\left|1,0,-1\right\rangle,\label{eq:1230} \\
\left|3_{+}\right\rangle  & \equiv & \left|v_{3}\right\rangle\otimes\left[\frac{1}{\sqrt{2}}\left(\left|1,1,0\right\rangle+\left|1,-1,0\right\rangle\right)\right],\nonumber \\
\left|3_{-}\right\rangle & \equiv & \left|v_{3}\right\rangle\otimes\left[\frac{1}{\sqrt{2}}\left(\left|1,1,0\right\rangle-\left|1,-1,0\right\rangle\right)\right].\nonumber
\end{eqnarray}
Correspondingly, we select the following three electromagnetic fields
\begin{eqnarray}
\emph{\textbf{E}}^{s}_{12}&=&\text{Re}\left\{\bm{e}^{s}_{-1}E^{s}_{12}e^{-\mathrm{i}\left(\omega_{12}t+\varphi_{12}\right)}\right\},\nonumber \\
\emph{\textbf{E}}^{s}_{23}&=&\text{Re}\left\{\bm{e}^{s}_{1} E^{s}_{23}e^{-\mathrm{i}\left(\omega_{23}t+\varphi_{23}\right)}\right\},\label{eq:Ejl}\\
\emph{\textbf{E}}^{s}_{13}&=&\text{Re}\left\{\bm{e}^{s}_{0} E^{s}_{13}e^{-\mathrm{i}\left(\omega_{13}t+\varphi_{13}\right)}\right\}.\nonumber
\end{eqnarray}
Because $\left|3_{+}\right\rangle$ and $\left|3_{-}\right\rangle$ are degenerate, two cyclic three-level substructures $\left|1\right\rangle \rightarrow \left|2\right\rangle \rightarrow \left|3_{+}\right\rangle \rightarrow \left|1\right\rangle$ and
$\left|1\right\rangle \rightarrow \left|2\right\rangle \rightarrow \left|3_{-}\right\rangle \rightarrow \left|1\right\rangle$ coexist at presence of the three electromagnetic fields. Thereby, the four-level model with two cyclic three-level substructures is formed (see Fig.~\ref{fig:model}) for the symmetric-top chiral molecule.

The corresponding Hamiltonian reads
\begin{eqnarray}
H
& = & \omega_{1}\left|1\right\rangle \left\langle 1\right|+\omega_{2}\left|2\right\rangle \left\langle 2\right|+\omega_{3}\left(\left|3_{+}\right\rangle \left\langle3_{+}\right|+\left|3_{-}\right\rangle \left\langle 3_{-}\right|\right) \nonumber \\
&   &
+\left[\Omega_{3_{+}2}e^{-\mathrm{i}\omega_{23}t}\left|3_{+}\right\rangle \left\langle 2\right|+\Omega_{3_{+}1}e^{-\mathrm{i}\omega_{13}t}\left|3_{+}\right\rangle \left\langle 1\right|  \right.  \nonumber \\
&   & +\Omega_{3_{-}2}e^{-\mathrm{i}\omega_{23}t}\left|3_{-}\right\rangle \left\langle 2\right|+\Omega_{3_{-}1}e^{-\mathrm{i}\omega_{13}t}\left|3_{-}\right\rangle \left\langle 1\right|  \label{eq:Hreal}  \\
&   & +\left. \Omega_{21}e^{-\mathrm{i}\omega_{12}t}\left|2\right\rangle  \left\langle 1\right|+\text{H.c.}\right]\nonumber
\end{eqnarray}
with the coupling strengths
\begin{equation}
\begin{aligned}
&\Omega_{21} = \frac{\sqrt{3}}{6}E^{s}_{12}e^{-\mathrm{i}\varphi_{12}}\left\langle v_{2}\right|\hat{\mu}^{m}_{z}\left|v_{1}\right\rangle, \\
&\Omega_{3_{+}2} = \frac{1}{4}E^{s}_{23}e^{-\mathrm{i}\varphi_{23}}\left\langle v_{3} \right|\hat{\mu}^{m}_{x}\left|v_{2}\right\rangle,\\
&\Omega_{3_{-}2} = \frac{\mathrm{i}}{4} E^{s}_{23}e^{-\mathrm{i}\varphi_{23}}\left\langle v_{3} \right|\hat{\mu}^{m}_{y}\left|v_{2}\right\rangle, \\
&\Omega_{3_{+}1} = \mathrm{i}\frac{\sqrt{3}}{6}  E^{s}_{13}e^{-\mathrm{i}\varphi_{13}}\left\langle v_{3} \right|\hat{\mu}^{m}_{y}\left|v_{1}\right\rangle,\\
&\Omega_{3_{-}1} = \frac{\sqrt{3}}{6} E^{s}_{13}e^{-\mathrm{i}\varphi_{13}}\left\langle v_{3} \right|\hat{\mu}^{m}_{x}\left|v_{1}\right\rangle.
\end{aligned}
\label{eq:Omegafinal}
\end{equation}
In each cyclic three-level substructure, the product of three transition electric dipoles in the molecule-fixed frame ($\left\langle v_{j}\right|\hat{\mu}_{\sigma^{\prime}}^{m}\left|v_{l}\right\rangle$) of the enantiomer differs in sign, which reflects the chirality of this model.
In addition, the transition electric dipoles in the molecule-fixed frame should be non-zero. Thus, we need to choose three suitable vibrational wave functions $\left|v_{l}\right\rangle\,(l=1,2,3)$ for the working states in Eq.~(\ref{eq:1230}).

\section{Vibrational wave function for  $\text{D}_{2}\text{S}_{2}$}\label{D2S2}
As follows, we take the molecule $\text{D}_{2}\text{S}_{2}$ as an example to introduce how to choose the vibrational wave functions of the working states. It must be ensured that all the transition electric dipoles in the molecule-fixed frame in Eq.~(\ref{eq:Omegafinal}) are non-zero and the product of three transition electric dipoles of each cyclic three-level substructure differs in sign with enantiomers of $\text{D}_{2}\text{S}_{2}$.

We are interested in the potential energy surface of the configuration space as a function of the following two coordinates introduced in Ref.~\cite{Shapiro4}. One coordinate is the dihedral angle $\tau$ between two $\text{DSS}$ planes, which describes the twist around the $\text{SS}$ bond. The other coordinate is the asymmetric $\text{S-D}$ stretching motion $\chi=R_{1}-R_{2}$ ($R_{1}$ and $R_{2}$ are the lengths of two $\text{S-D}$ bonds respectively). The potential energy surfaces for $\tau$ and $\chi$ are in the form of a two-dimensional double well. The coordinate $\tau$ reflects the chirality. The left-handed states (expressed by the superscript ``$L$'') are localized on the left-handed part ($0\leq\tau<\pi$) of the potential energy surface and the right-handed states (expressed by the superscript ``$R$'') are localized on the right-handed part ($\pi\leq\tau<2\pi$). As follows, we only consider the chiral states.

We use $\left|\tilde{m}\right\rangle^{Q}_{\tau}$ and $\left|\tilde{n}^{\pm}\right\rangle_{\chi}$ to represent the eigenstates for the freedoms $\tau$ and $\chi$, respectively. Here $Q=L,R$ represents the chirality and is related to the range of $\tau$. And $\tilde{m}$ and $\tilde{n}$ are the non-negative integers and increase in unit steps. $\left|\tilde{n}^{+}\right\rangle_{\chi}$ ($\left|\tilde{n}^{-}\right\rangle_{\chi}$) refers to the state with even (odd) parity for $\chi$. Moreover, it is reasonable to assume that the eigenstates $\left|\tilde{m}\right\rangle^{Q}_{\tau}$ and $\left|\tilde{n}^{\pm}\right\rangle_{\chi}$ are dynamically decoupled~\cite{Shapiro4}, then the vibrational wave functions can be written as a product of $\left|\tilde{m}\right\rangle^{Q}_{\tau}$ and $\left|\tilde{n}^{\pm}\right\rangle_{\chi}$, i.e., $\left|v\right\rangle_{Q}\equiv\left|\tilde{m}\right\rangle^{Q}_{\tau}\otimes\left|\tilde{n}^{\pm}\right\rangle_{\chi}$.

The transition electric dipole in the molecule-fixed frame in Eq.~(\ref{eq:Omegafinal}) can be demonstrated as the integral over the surface of configuration space
\begin{eqnarray}
&&  {}_{\chi}\left\langle\tilde{n}^{\pm} \right| \otimes{}^{Q}_{\tau}\left\langle \tilde{m}\right|
\hat{\mu}_{\sigma^{\prime}}^{m}
\left|\tilde{m}^{\prime}\right\rangle^{Q}_{\tau}\otimes\left|\tilde{n}^{\prime\pm}\right\rangle_{\chi}\label{eq:v}\\
&&=\iint \psi^{Q\ast}_{\tilde{m},\tilde{n}^{\pm}}\left(\tau,\chi\right)
\mu_{\sigma^{\prime}}^{m}\left(\tau,\chi\right)
\psi^{Q}_{\tilde{m}^{\prime},\tilde{n}^{\prime\pm}} \left(\tau,\chi\right) d\tau d\chi\nonumber
\end{eqnarray}
with $\mu_{\sigma^{\prime}}^{m}\left(\tau,\chi\right)=\left\langle\tau,\chi\right|\hat{\mu}_{\sigma^{\prime}}^{m}\left|\tau,\chi\right\rangle$ being the component of $\hat{\mu}_{\sigma^{\prime}}^{m}$ in the representation of $\left|\tau,\chi\right\rangle$ and $\psi_{\tilde{m},\tilde{n}^{\pm}}^{Q}\left(\tau,\chi\right)=\left\langle\tau,\chi|\tilde{m}\right\rangle^{Q}_{\tau}\otimes\left|\tilde{n}^{\pm}\right\rangle_{\chi}$ being the wave function of the state $\left|\tilde{m}\right\rangle^{Q}_{\tau}\otimes\left|\tilde{n}^{\pm}\right\rangle_{\chi}$ in the representation of $\left|\tau,\chi\right\rangle$.

The symmetrical relationships of $\mu_{\sigma^{\prime}}^{m}\left(\tau,\chi\right)$ and $\psi_{\tilde{m},\tilde{n}^{\pm}}^{Q}\left(\tau,\chi\right)$ with respect to the coordinate $\chi$ are~\cite{Shapiro4}
\begin{eqnarray}
\mu_{z}^{m}\left(\tau,\chi\right)    &=&  \mu_{z}^{m}\left(\tau,-\chi\right),\nonumber \\
\mu_{x,y}^{m}\left(\tau,\chi\right)  &=& -\mu_{x,y}^{m}\left(\tau,-\chi\right), \label{eq:dipole}\\
\psi_{\tilde{m},\tilde{n}^{+}}^{Q}\left(\tau,\chi\right)  &=& \psi_{\tilde{m},\tilde{n}^{+}}^{Q}\left(\tau,-\chi\right), \nonumber\\
\psi_{\tilde{m},\tilde{n}^{-}}^{Q}\left(\tau,\chi\right)  &=& -\psi_{\tilde{m},\tilde{n}^{-}}^{Q}\left(\tau,-\chi\right). \nonumber
\end{eqnarray}
Due to the symmetrical relationships in Eq.~(\ref{eq:dipole}), the selection of $\left|\tilde{n}^{+}\right\rangle_{\chi}$ or $\left|\tilde{n}^{-}\right\rangle_{\chi}$ plays a key role in determining whether the transition electric dipoles in the molecule-fixed frame $\left\langle v_{j}\right|\hat{\mu}_{\sigma^{\prime}}^{m}\left|v_{l}\right\rangle$ in Eq.~(\ref{eq:Omegafinal}) are zero. All the non-zero transition electric dipoles in the molecule-fixed frame are
\begin{eqnarray}
{}_{\chi}\left\langle\tilde{n}^{+}\right|\otimes{}_{\tau}^{Q}\left\langle \tilde{m}\right|
\hat{\mu}_{z}^{m}
\left|\tilde{m}^{\prime}\right\rangle^{Q}_{\tau}\otimes\left|\tilde{n}^{\prime+}\right\rangle_{\chi} & \neq & 0,\nonumber\\
{}_{\chi}\left\langle\tilde{n}^{-}\right|\otimes{}_{\tau}^{Q}\left\langle \tilde{m}\right|
\hat{\mu}_{z}^{m}
\left|\tilde{m}^{\prime}\right\rangle^{Q}_{\tau}\otimes\left|\tilde{n}^{\prime-}\right\rangle_{\chi} & \neq & 0,\label{eq:n}\\
{}_{\chi}\left\langle\tilde{n}^{+}\right|\otimes{}_{\tau}^{Q}\left\langle \tilde{m}\right|
\hat{\mu}_{x,y}^{m}
\left|\tilde{m}^{\prime}\right\rangle^{Q}_{\tau}\otimes\left|\tilde{n}^{\prime-}\right\rangle_{\chi} & \neq & 0.\nonumber
\end{eqnarray}
Thus, a suitable set of the vibrational wave functions for the working states can be selected, such as, $\left|v_{1}\right\rangle_{Q}=\left|0\right\rangle^{Q}_{\tau} \otimes \left|0^{+}\right\rangle_{\chi}$, $\left|v_{2}\right\rangle_{Q}=\left|1\right\rangle^{Q}_{\tau} \otimes \left|0^{+}\right\rangle_{\chi}$, and $\left|v_{3}\right\rangle_{Q}=\left|1\right\rangle^{Q}_{\tau} \otimes \left|0^{-}\right\rangle_{\chi}$.

The symmetrical relationships of $\mu_{\sigma^{\prime}}^{m}\left(\tau,\chi\right)$ and $\psi_{\tilde{m},\tilde{n}^{\pm}}^{Q}\left(\tau,\chi\right)$ with respect to $\tau$ are
\begin{eqnarray}
\mu_{z}^{m}\left(\pi-\tau,\chi\right)    &=& -\mu_{z}^{m}\left(\pi+\tau,\chi\right),\nonumber \\
\mu_{x,y}^{m}\left(\pi-\tau,\chi\right)  &=&  \mu_{x,y}^{m}\left(\pi+\tau,\chi\right),\label{eq:dipole2}\\
\psi_{\tilde{m},\tilde{n}^{+}}^{L}\left(\pi-\tau,\chi\right) &=& \psi_{\tilde{m},\tilde{n}^{+}}^{R}\left(\pi+\tau,\chi\right), \nonumber
\\
\psi_{\tilde{m},\tilde{n}^{-}}^{L}\left(\pi-\tau,\chi\right) &=& \psi_{\tilde{m},\tilde{n}^{-}}^{R}\left(\pi+\tau,\chi\right). \nonumber
\end{eqnarray}
The symmetrical relationships in Eq.~(\ref{eq:dipole2}) determine the chirality dependency of the transition electric dipole in the molecule-fixed frame in Eq.~(\ref{eq:Omegafinal}), that is
\begin{eqnarray}
_{L}\left\langle v_{2}\right|\hat{\mu}_{z}^{m} \left|v_{1}\right\rangle_{L}  &=&
-_{R}\left\langle v_{2}\right|\hat{\mu}_{z}^{m}\left|v_{1}\right\rangle_{R},\nonumber \\
_{L}\left\langle v_{3}\right|\hat{\mu}_{x,y}^{m}  \left|v_{1}\right\rangle_{L} &=&
 _{R}\left\langle v_{3}\right|\hat{\mu}_{x,y}^{m} \left|v_{1}\right\rangle_{R}, \\
_{L}\left\langle v_{3}\right|\hat{\mu}_{x,y}^{m} \left|v_{2}\right\rangle_{L} &=&
_{R}\left\langle v_{3}\right|\hat{\mu}_{x,y}^{m} \left|v_{2}\right\rangle_{R}. \nonumber
\end{eqnarray}
Therefore, the chirality dependence of chiral molecules reflects in the coupling strengths in Eq.~(\ref{eq:Hreal}), that is
\begin{equation}
\begin{aligned}
& \Omega_{3_{+}1}^{L}=\Omega_{3_{+}1}^{R}=\Omega_{3_{+}1},
~~\Omega_{3_{-}1}^{L}=\Omega_{3_{-}1}^{R}=\Omega_{3_{-}1},\\
& \Omega_{3_{+}2}^{L}=\Omega_{3_{+}2}^{R}=\Omega_{3_{+}2},
~~\Omega_{3_{-}2}^{L}=\Omega_{3_{-}2}^{R}=\Omega_{3_{-}2},\\
& \Omega_{21}^{L}    =-\Omega_{21}^{R}   =\Omega_{21}.
\label{eq:OmegaLR}
\end{aligned}
\end{equation}
Here, we have added the superscript ``$L$''  or ``$R$'' to denote the left-handed or right-handed chiral molecule. In the following, when referring to left-handed (right-handed) chiral molecules, we will add the superscript. When there is no superscript, we refer to general molecules. Hence, when the enantiomers couple with the same electric fields, the related coupling strengths $\pm\Omega_{21}$ reflect the chiral difference.

\section{Enantio-specific state transfer} \label{V}
In the above four-level model described by Eq.~(\ref{eq:Hreal}) (see Fig.~\ref{fig:model}), the one-photon process $\left|1\right\rangle \rightarrow \left|2\right\rangle$  is chirality-dependent, while the two-photon process $\left|1\right\rangle \rightarrow \left|3_{+}\right\rangle \rightarrow \left|2\right\rangle$ or $\left|1\right\rangle \rightarrow \left|3_{-}\right\rangle \rightarrow \left|2\right\rangle$ is chirality-independent. Then the interference between the one-photon process and the two-photon process in the cyclic three-level substructure is chirality-dependent. Such chirality-dependent interferences give rise to the chirality-dependent dynamics for the enantiomer to achieve the enantio-specific state transfer (or the enantio-detection and enantio-separation).

Since the dynamics of the four-level model for symmetric-top chiral molecules is different from that of the cyclic three-level model, the schemes~\cite{Kral2,Kral,Li3,Jia2,Ye1,Torosov1,Torosov2,Wu} of the enantio-specific state transfer based on the cyclic three-level models may be no longer applicable. Based on the four-level model, we will introduce two dynamic schemes to achieve the enantio-specific state transfer for symmetric-top chiral molecules.

Under the three-photon resonance condition $\omega_{12}+\omega_{23}-\omega_{13}=0$, Hamiltonian~(\ref{eq:Hreal}) can be rewritten in the time-independent form in the interaction picture as
\begin{eqnarray}
H^{\prime}
& = & \Delta_{12}\left|2\right\rangle \left\langle 2\right|
     +\Delta_{13}\left(\left|3_{+}\right\rangle \left\langle 3_{+}\right|+\left|3_{-}\right\rangle \left\langle 3_{-}\right| \right) \nonumber  \\
&   &+\left[\Omega_{21}\left|2\right\rangle \left\langle 1\right|
           +\Omega_{3_{+}2}\left|3_{+}\right\rangle \left\langle 2\right|
           +\Omega_{3_{+}1}\left|3_{+}\right\rangle \left\langle 1\right| \right. \nonumber \\
&   &+\left.\Omega_{3_{-}2}\left|3_{-}\right\rangle \left\langle 2\right|
           +\Omega_{3_{-}1}\left|3_{-}\right\rangle \left\langle 1\right| +\text{H.c.}\right]
\label{eq:Hinnprime}
\end{eqnarray}
with the detunings $\Delta_{lj} \equiv \left(\omega_{j}-\omega_{l}\right)-\omega_{lj}$.

\begin{figure}[htbp]
\centering
\includegraphics[width=0.80\linewidth]{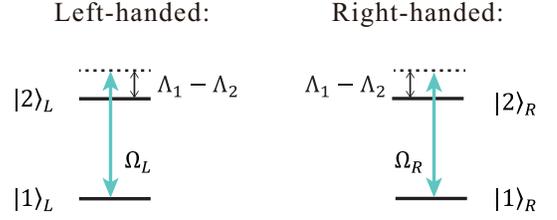}
\caption{The effective two-level model with the same effective detunings $\Lambda_{1}-\Lambda_{2}$ and the chirality-dependent effective couplings $\Omega_{L}=\Omega_{\text{eff}}+\Omega_{21}$ for the left-handed chiral molecules and $\Omega_{R}=\Omega_{\text{eff}}-\Omega_{21}$ for the right-handed chiral molecules.
\label{fig:twolevel}}
\end{figure}

We now consider the case that the electromagnetic field $\emph{\textbf{E}}^{s}_{12}$ is resonant with the transition $\left|1\right\rangle \leftrightarrow \left|2\right\rangle$, i.e., $\Delta_{12}=0$. Then the other two electromagnetic fields $\emph{\textbf{E}}^{s}_{23}$ and $\emph{\textbf{E}}^{s}_{13}$ are in two-photon resonance, i.e., $\Delta_{23}=\Delta_{13}=\Delta$. Therefore, the enantio-specific state transfer can be realized by adjusting the electromagnetic fields (the field amplitudes and the phases) and the detuning.

\subsection{Effective two-level model}
In what follows, for the sake of simplicity, we assume the large-detuning condition
\begin{eqnarray}
\left|\Delta\right| & \gg & \left|\Omega_{3_{+}1}\right|\sim\left|\Omega_{3_{+}2}\right|\gg\left|\Omega_{21}\right|,  \nonumber \\
\left|\Delta\right| & \gg & \left|\Omega_{3_{-}1}\right|\sim\left|\Omega_{3_{-}2}\right|\gg\left|\Omega_{21}\right|,
\label{eq:detuning}
\end{eqnarray}
so that the four-level model can reduce to an effective two-level one by eliminating adiabatically the excited levels $\left|3_{+}\right\rangle$ and $\left|3_{-}\right\rangle$. To this end, we decompose the  Hamiltonian as $H^{\prime} =H_{0}+H_{1}+H_{2}$ with the zeroth-order Hamiltonian
$H_{0}=\Delta (\left|3_{+}\right\rangle \left\langle 3_{+}\right|
+\left|3_{-}\right\rangle \left\langle 3_{-}\right|)$, the first-order term
$H_{1}=\Omega_{3_{+}2}\left|3_{+}\right\rangle \left\langle 2\right|
+      \Omega_{3_{+}1}\left|3_{+}\right\rangle \left\langle 1\right|+
       \Omega_{3_{-}2}\left|3_{-}\right\rangle \left\langle 2\right|
+      \Omega_{3_{-}1}\left|3_{-}\right\rangle \left\langle 1\right|+\text{H.c.}$, and the second-order term
$H_{2}=\Omega_{21}\left|2\right\rangle \left\langle 1\right|+\text{H.c.}$. By the anti-Hermitian operator $S=(\Omega_{3_{+}2}\left|3_{+}\right\rangle \left\langle 2\right|+\Omega_{3_{+}1}\left|3_{+}\right\rangle \left\langle 1\right|+\Omega_{3_{-}2}\left|3_{-}\right\rangle \left\langle 2\right|+\Omega_{3_{-}1}\left|3_{-}\right\rangle \left\langle 1\right|-\text{H.c.})/\Delta$, the unitary transformation
$H_{\text{eff}}=\exp{\left(-S\right)} H^{\prime} \exp{\left(S\right)}\simeq H_{0}+\left[H_{1},S\right]/2+H_{2}$ defines the following effective Hamiltonian
\begin{equation}
\begin{aligned}
H_{\text{eff}} = &
\Lambda_{1} \left|1\right\rangle \left\langle 1\right|
+\Lambda_{2} \left|2\right\rangle \left\langle 2\right|
+\left[\left(\Omega_{\text{eff}}+\Omega_{21}\right) \left|2\right\rangle \left\langle 1\right|+\text{H.c.}\right]\\
+&\Delta \left(\left|3_{+}\right\rangle \left\langle3_{+}\right|
+\left|3_{-}\right\rangle \left\langle 3_{-}\right|\right)
+\left(\Omega^{\prime}_{\text{eff}}\left|3_{+}\right\rangle \left\langle 3_{-} \right|+\text{H.c.}\right)
\end{aligned}
\label{eq:Heffinn1}
\end{equation}
with the energy shifts $\Lambda_{1}= -(|\Omega_{3_{+}1}|^{2}+|\Omega_{3_{-}1}|^{2})/\Delta$ and $\Lambda_{2}=-(|\Omega_{3_{+}2}|^{2}+|\Omega_{3_{-}2}|^{2})/\Delta$, the effective coupling $\Omega_{\text{eff}}=-(\Omega_{3_{+}1}\Omega^{\ast}_{3_{+}2}+\Omega_{3_{-}1}\Omega^{\ast}_{3_{-}2})/\Delta$ and $\Omega_{\text{eff}}^{\prime}=(\Omega^{\ast}_{3_{+}1}\Omega_{3_{-}1}+\Omega^{\ast}_{3_{+}2}\Omega_{3_{-}2})/\Delta$. In addition, the non-zero $\Omega_{\text{eff}}$ requires the vibrational wave functions of the working state to satisfy the condition of $\left|v_{1}\right\rangle \neq \left|v_{2}\right\rangle$.

\begin{figure}[htbp]
\centering
\includegraphics[width=0.90\linewidth]{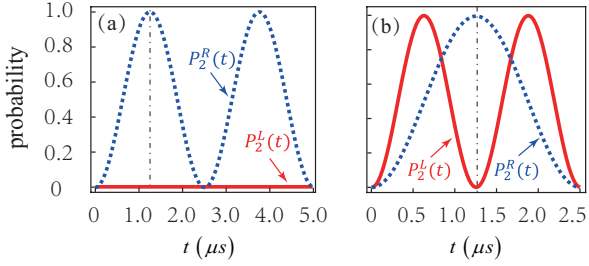}
\caption{Dynamic ways of the enantio-specific state transfer for symmetric-top chiral molecules based on Eq.~(\ref{eq:Heff12}) for (a) the first way with $\Omega_{21}=-\Omega_{\text{eff}}=-2\pi\times0.10\,\text{MHz}$ and (b) the second way with $\Omega_{21}=3\Omega_{\text{eff}}=2\pi\times0.30\,\text{MHz}$. The probability occupying the state $\left|2\right\rangle_{L}$ ($\left|2\right\rangle_{R}$) of the left-handed (right-handed) chiral molecules, $P^{L}_{2}(t)$  [$P^{R}_{2}(t)$], is denoted by the red solid (blue dashed) line.
\label{fig:inner}}
\end{figure}

In Eq.~(\ref{eq:Heffinn1}), the states $\left|3_{+}\right\rangle$ and $\left|3_{-}\right\rangle$ are decoupled from the states $\left|1\right\rangle$ and $\left|2\right\rangle$. That means we get an effective two-level model in the subspace $\left\{\left|1\right\rangle,\,\left|2\right\rangle\right\}$. Then, according to the chirality dependency of chiral molecules in Eq.~(\ref{eq:OmegaLR}), the effective Hamiltonian for the enantiomer in the subspace $\{\left|1\right\rangle_{Q},\,\left|2\right\rangle_{Q}\}$ $(Q=L,R)$ reads
\begin{eqnarray}
H^{\text{eff}}_{12,Q} &=&
\frac{1}{2}(\Lambda_{1}-\Lambda_{2})
(\left|1\right\rangle_{QQ}\left\langle 1\right|
-\left|2\right\rangle_{QQ} \left\langle 2\right|)\nonumber\\
&&+(\Omega_{Q} \left|2\right\rangle_{QQ} \left\langle 1\right|+\text{H.c.})
\label{eq:Heff12}
\end{eqnarray}
with the same effective detunings $\Lambda_{1}-\Lambda_{2}$ and chirality-dependent effective couplings $\Omega_{L}=\Omega_{\text{eff}}+\Omega_{21}$ $(\Omega_{R}=\Omega_{\text{eff}}-\Omega_{21})$ for the left-handed (right-handed) chiral molecules (see Fig.~\ref{fig:twolevel}). Here, we  have omitted the item $(\Lambda_{1}+\Lambda_{2})(\left|1\right\rangle_{QQ}\left\langle 1\right|+\left|2\right\rangle_{QQ} \left\langle 2\right|)/2$ for simplicity.

\subsection{Dynamics of the enantio-specific state transfer}
Assumed the initial state as $\left|\Psi(0)\right\rangle_{Q}=\left|1\right\rangle_{Q}$, the possibility excited into state $\left|2\right\rangle_{Q}$ can be solved analytically according to the Schr{\"o}dinger equation $\mathrm{i} \partial_{t} \left|\Psi(t)\right\rangle_{Q}=H^{\text{eff}}_{12,Q}\left|\Psi(t)\right\rangle_{Q}$ as
\begin{equation}
P_{2}^{Q}(t)=\left|\frac{\Omega_{Q}}{\widetilde{\Omega}_{Q}}\right|^{2}
\sin^{2}(\widetilde{\Omega}_{Q} t).
\label{eq:p2}
\end{equation}
Here $\widetilde{\Omega}_{Q}=\sqrt{\left|\Omega_{Q}\right|^{2}+\left(\Lambda_{1}-\Lambda_{2}\right)^{2}/4}$, and $2\widetilde{\Omega}_{Q}$ is the Rabi oscillation frequency of the probability occupying the state $\left|2\right\rangle_{Q}$. The corresponding Rabi oscillation period is $T_{Q} = \pi/\widetilde{\Omega}_{Q}$.

By adjusting the effective detunings and the chirality-dependent effective couplings, the enantiomers prepared initially in the same-energy levels can be evolved to different-energy levels, i.e., the achievement of the enantio-specific state transfer. In what follows, we will show two dynamic ways of the enantio-specific state transfer. For simplicity, we will choose $\Lambda_{1}=\Lambda_{2}$. That means $\widetilde{\Omega}_{Q}=|\Omega_{Q}|$ and $T_{Q} = \pi/|\Omega_{Q}|$.

For the first way, the left-handed chiral molecules always remain in the initial state $\left|1\right\rangle_{L}$ while the right-handed chiral molecules experience a half-integer period of its Rabi oscillation, then the perfect enantio-specific state transfer is achieved at $t=(N+1/2)T_{R}$ with $N$ natural number. For this purpose, the three electromagnetic fields should be appropriately adjusted so that
\begin{equation}
\Omega_{21}=-\Omega_{\text{eff}}.
\label{eq:wLR1}
\end{equation}
In Fig.~\ref{fig:inner}(a), an example of the first way is illustrated with the parameters  $\Omega_{21}=-\Omega_{\text{eff}}=-2\pi\times0.10\,\text{MHz}$ based on Eq.~(\ref{eq:Heff12}). It shows that the perfect enantio-specific state transfer is achieved at $t=1.25\,\mu\text{s}$.

Similarly, the right-handed chiral molecules always remain in the initial state $\left|1\right\rangle_{R}$ while the left-handed chiral molecules experience a half-integer period of its Rabi oscillation. At $t=(N+1/2)T_{L}$, the perfect enantio-specific state transfer is realized.

For the second way, when the left- and right-handed chiral molecules simultaneously experience half-integer and integer periods of their corresponding Rabi oscillations, i.e.,
\begin{equation}
N_{L}T_{L}=\left(N_{R}+\frac{1}{2}\right)T_{R}
\label{eq:nLR}
\end{equation}
with $N_{L,R}$ natural number, the perfect enantio-specific state transfer is achieved at $t=N_{L}T_{L}$. For this purpose, the three electromagnetic fields should be appropriately adjusted. Taking the case of $\Omega_{21}>\Omega_{\text{eff}}>0$ as an example, we could adjust the three electromagnetic fields to meet
\begin{equation}
\Omega_{21} = \frac{2N_{L}+2N_{R}+1}{2N_{L}-2N_{R}-1}\Omega_{\text{eff}}.
\label{eq:wLR2}
\end{equation}
In Fig.~\ref{fig:inner}(b), an example of the second way is illustrated with the parameters $\Omega_{21}=3\Omega_{\text{eff}}=2\pi\times0.30\,\text{MHz}$ based on Eq.~(\ref{eq:Heff12}). It shows that the perfect enantio-specific state transfer is achieved at $t=1.25\,\mu\text{s}$.

Similarly, when $\left(N_{L}+1/2\right)T_{L}=N_{R}T_{R}$, the left-handed chiral molecules experience half-integer periods and simultaneously the right-handed chiral molecules experience integer periods of its Rabi oscillation. Then the perfect enantio-specific state transfer is realized at $t=N_{R}T_{R}$.

The above two ways of the perfect enantio-specific state transfer are based on Eq.~(\ref{eq:Heff12}), which is obtained by adiabatically eliminating the excited states $\left|3_{+}\right\rangle$ and $\left|3_{-}\right\rangle$ in the original Hamiltonian~(\ref{eq:Hinnprime}). Thus, these two dynamic ways to achieve the enantio-specific state transfer for symmetric-top chiral molecules are approximately perfect.

\section{Conclusion} \label{Conclusion}
In conclusion, according to the electric-dipole selection rules of gaseous symmetric-top chiral molecules, we have selected the appropriate working states and the corresponding three electromagnetic fields to realize the chirality-dependent four-level model with two cyclic three-level substructures. Under the large-detuning condition, we can reduce the four-level model to the effective two-level one with the same effective detuning but the chirality-dependent effective couplings. Then we have used two dynamic ways to achieve the approximately perfect enantio-specific state transfer for symmetric-top chiral molecules.
Comparing with previous researches of enantio-specific state transfer for asymmetric-top chiral molecules based on cyclic three-level systems, our work of four-level model provides an important complement to enantio-specific state transfer of (approximate) symmetric-top chiral molecules where the cyclic three-level systems do not exist.  In addition, the researches of the enantio-detection, enantio-separation, and enantio-conversion~\cite{Shapiro,Salam,Fujimuraa,Hirota,Lehmann4,Jia1,Ye2,Chen,Xu,Kang,Chen2,Li2,ShapiroLiX,Kral2,Kral,Li3,Jia2,Ye1,Torosov1,Torosov2,Wu} of chiral molecules are always meaningful and challenging tasks. Our four-level model will play important roles in the future investigations of these issues for symmetric-top chiral molecules.

\section{Acknowledgement}
This work was supported by the Natural Science Foundation of China (Grants No. 12074030, No. 11774024, No. 12088101, No. U1930402, and No. U1930403) and the China Postdoctoral Science Foundation (Grant No. 2021M690323).


\end{document}